\documentclass[prc,twocolumn,showpacs,floatfix,nofootinbib,preprintnumbers,superscriptaddress,amsmath,amssymb]{revtex4-1}

\usepackage{graphicx}  
\usepackage{hyperref}

\usepackage{epsfig}  
\usepackage{bm}  
\usepackage{color}  
\usepackage{float}  
\usepackage{dcolumn}   
\usepackage{multirow}   
\usepackage{longtable}  
\def\be{\begin{equation}}  
\def\ee{\end{equation}}

\begin{document}

\title{Benchmarking Nuclear Fission Theory}  
  
\author{G.F. Bertsch}  
\affiliation{Department of Physics and Institute for Nuclear Theory,  
University of Washington, Seattle, Washington 98195, USA}  
  
\author{W. Loveland}  
\affiliation{Department of Chemistry, Oregon State University, Corvallis, Oregon 97331, USA}  
  
\author{W. Nazarewicz}  
 \affiliation{Department of Physics and Astronomy and NSCL/FRIB Laboratory, 
Michigan State University, East Lansing, Michigan  48824, USA\\ 
Physics Division, Oak Ridge National Laboratory, Oak Ridge, Tennessee 37831, USA} 
 
\author{P. Talou}  
\affiliation{Nuclear Physics Group, Theoretical Division, Los Alamos National Laboratory,  
Los Alamos, New Mexico 87545, USA}  
  
\begin{abstract}  
We suggest a small set of fission observables to be used as test cases  
for validation of theoretical calculations.  The purpose is to provide common data to   
facilitate the comparison of   
different fission theories and models.  The proposed observables  
are chosen from fission barriers, spontaneous fission lifetimes,   
fission yield characteristics, and fission isomer excitation energies.  
\end{abstract}  
  
\maketitle  
 
\section{Motivation}  
  
Nuclear fission is a very complex process and its theory presents an enormous challenge.    
As Bohr and Wheeler stated in their 1939 pioneering paper \cite{Boh39}, theoretical progress in the theory of fission would in all likelihood take time to resolve: ``An accurate estimate for the stability of a heavy nucleus against fission in its ground state will, of course, involve a very complicated mathematical problem". Indeed, even in the present era of  
extensive computer resources, a comprehensive microscopic explanation of nuclear fission rooted in interactions between protons and neutrons still eludes us. Consequently,  
it remains difficult for both  
experimentalists and theorists to assess various models of fission 
and their predictions. To address this situation, it would be  
very useful if different theoretical approaches could be   
easily compared.  Most  
importantly, such reporting should promote a closer interaction  
between theorists and experimentalists to stimulate new  
experiments that can differentiate between models or unveil  
new phenomena.

To that end, we would like to  
suggest a list of experimental observables, or evaluated empirical quantities,  that are  well established,  
and could serve as benchmarks of the accuracy of a theoretical  
approach.  Our recommendation for future model development work is to present along with predictions  
of  
a theory, the results when applied to this small set of   
data.  The benchmark cases we have selected are basic fission  
observables in nuclei that are  well known  
experimentally.  The observables in the benchmark are: fission  
barriers, fission mass distributions, total kinetic energies of fission  
fragments, spontaneous fission  
lifetimes, and  fission isomer excitation energies.  This  
leaves out a rich variety of interesting phenomena that includes  
kinetic energy distributions of  fission yields, scission neutrons, and barrier state spectroscopy.  The theory for these  
quantities is not as well developed.  Hopefully, candidate  
theories for the more complex phenomena would be sufficiently general  to apply them to 
the basic benchmarks.  
  
It is also important that the results be reported in a way that makes comparisons easy.  In particular, we would like to know how the theory performs on average for the data set, if the parameters of the theory have not been adjusted to the data.  We would also like to know how well the theory describes the fluctuations of individual data.    
  
We understand that a large community of experimentalists, theorists, and evaluators has been working for a long time on developing standards and benchmarks related to fission data. The purpose of the present contribution is not to reproduce, or even attempt to reproduce, this large body of work, but instead to select from it a subset of well-known fission data that can be readily used by fission theorists to guide and test their work. 
  
When dealing with fission data, it is important to realize that what is considered ``experimental data" is often the result of a more or less complicated deconvolution process related to a physical observable. This caveat will be repeated and illustrated wherever it applies. 
 
Finally,  as the purpose of  these notes is to stimulate   benchmarking   rather than provide critical evaluation of various models of fission, we choose not to provide specific examples of theoretical calculations. Here, we would like to draw the reader's attention to the talks presented at the INT Program 13-3, posted online \cite{INT}, which contain a wealth of useful information about the current status of fission theory. 
  
\section{The Benchmarks}  
 
\subsection{Fission Barrier Heights}  
  
\begin{table}[hbt]  
\caption{  
Fission barrier parameters for the even-even actinides.  $E_A$ and  $E_B$ are the empirical heights of the inner and outer fission barrier,  
respectively \cite{Cap09a}.  The  
uncertainty on the empirical barrier heights ranges from $0.3$\,MeV \cite{Smi93} to $1$\,MeV.}  
\label{tab:2}        
\begin{ruledtabular}  
\begin{tabular}{ccccc}  
Z & A & Symbol &  $E_A$ (MeV) & $E_B$ (MeV)\\  
\hline  \\[-8pt]  
90 &	230 & Th &  6.1 &  6.8\\  
90&232&	Th&		5.8&	6.7\\  
92&	232&	U&		4.9&		5.4\\  
92&	234&	U&		4.8&		5.5\\  
92&	236&	U&		5&		5.67\\  
92&	238&	U&		6.3&		5.5\\  
94&	238&	Pu&		5.6&		5.1\\  
94&	240&	Pu& 	6.05&		5.15\\  
94&	242&	Pu&		5.85&		5.05\\  
94&	244&	Pu&	    5.7&		4.85\\  
96&	241&	Cm&		7.15&		5.5\\  
96&	242&	Cm&	    6.65&		5\\  
96&	244&	Cm&	    6.18&		5.1\\  
96&	246&	Cm&		6   &		4.8\\  
96&	248&	Cm&	    5.8&		4.8  
\end{tabular}  
\end{ruledtabular}  
\end{table}  
  
The concept of a fission barrier height is fraught with ambiguity \cite{INT}.  A theoretical definition is the energy difference between the ground state and the highest saddle point in a shape-constrained potential energy surface (PES) that has the lowest energy for all possible paths leading to fission from the ground state. If the theory treats the angular momentum of the nucleus, the benchmark should be for the PES corresponding to the angular momentum of the ground state.  We have chosen 15 examples for the benchmarks, including the well-known nuclei for reactor physics, and some examples with isotope chains ranging from $Z=90$ to $Z=96$ and an example beyond Pu to better exhibit the $Z$-dependence of the barriers.  The empirical barriers are taken from RIPL-3 compilation \cite{Cap09a}.  
  
Contrary to cross sections, fission barriers are not physical observables, and ``empirical" barriers are inferred from measured cross sections using particular models for the PES, the collective inertia tensor, and the level density on top of the barrier. The presence of a double-humped, or more complicated,  structure along the predicted fission pathways further complicate matters  as significant deviations from the traditional Hauser-Feshbach calculations of fission probabilities have to be considered. 
  
The study in Ref.~\cite{Smi93} concludes that fission barrier heights can be known to about $\pm$0.3 MeV, with only little sensitivity to the particular prescription chosen for describing the level density on top of the barrier. We should consider this uncertainty as a lower limit, since complications may arise with a more realistic treatment of  penetrabilities associated with complicated pathways.  
 
In addition to providing benchmark values in Table~\ref{tab:2} against which theoretical calculations can be compared, trends in inner and outer fission barrier heights as a function of mass number and fissility parameter $Z^2/A$ should also serve as a guide. For lower-$Z$ actinides, e.g., Th isotopes, inner barrier heights are lower than outer barrier heights. This trend is reversed for heavier actinides, e.g., Cm isotopes. 
 
\subsection{Fission Isomer Excitation Energies}  
  
One of the most challenging aspects of fission theory is to correctly predict the energies and half-lives of the superdeformed intermediate states of the fissioning nucleus, the spontaneously fissioning shape isomers.  The excitation energies are typically 2-3 MeV  in the second minimum of the fission barrier.  Spectroscopic studies of the transitions between the states in the second minimum have shown that the moments of inertia associated with the rotational bands are those expected for nuclei with an axes ratio of 2:1 -- a result confirmed by studies of the quadrupole moments \cite{Metag}.  All of these facts represent a significant constraint on, and a challenge for, fission theories.  
 
An isomer excitation energy can be obtained by analyzing experimental data on the excitation energy dependence of the cross sections for formation of the isomer, and in particular near the threshold of the rising curves. Most of these experimental data come from neutron evaporation and particle transfer reactions. As for fission barrier heights, the inferred isomer energy is model-dependent, and has to be considered carefully. 
  
As discussed in Ref.~\cite{Bjo80}, the analysis of the experimental excitation curves is easier in the case of fissioning doubly-odd nuclei, where simplifying assumptions can be made on the level density representation used in the cross section calculations. Even in those cases, however, the uncertainty on the isomer energy is probably at least equal to the uncertainty ($\sim$0.3-1.0 MeV) on fission barrier heights, as discussed above.

\begin{table}[hbt]  
\caption{Table of (even-even) Fission isomer excitation  energies $E_{\rm II}$ \cite{Sin02,NNDC}}  
\label{tab:3}       
\begin{ruledtabular}  
\begin{tabular}{ccc}  
Nuclide & $E_{\rm II}$ (keV) & $T_{1/2}$\\  
\hline  \\[-8pt]  
$^{236}$U & 2750 & 120 ns \\  
$^{238}$U & 2557.9 & 280 ns \\  
$^{238}$Pu & $\sim$ 2400  & 0.6 ns \\  
$^{240}$Pu & $\sim$ 2800  & 3.7 ns \\  
$^{242}$Pu & $\sim$ 2000 & 28 ns \\  
           & 2000 & 3.5 ns \\  
$^{240}$Cm & $\sim$ 3000 & 55 ns \\  
$^{242}$Cm & $\sim$1900 & 40 ps \\  
$^{244}$Cm & $\sim$ 2200 & $\le$5 ps 
\end{tabular}  
\end{ruledtabular}  
\end{table}  
  
\subsection{Spontaneous Fission Lifetimes}  
   
The examples chosen in Table~\ref{tab:1} are for illustrative purposes only. Many more spontaneous fission half-lives have been measured and analyzed, as reported in Ref.~\cite{Hol00}. For the examples we have chosen the well-known $^{240}$Pu lifetime together with two cases among heavier actinide elements that exhibit extreme variations in lifetimes.  
 
\begin{table}[htb]  
\caption{Spontaneous fission half-lives \cite{Hol00}.}  
\label{tab:1}       
\begin{ruledtabular}  
\begin{tabular}{cr}  
Nuclide & $T_{\rm SF}$\\  
\hline  \\[-8pt]  
$^{240}$Pu&1.14 $\pm$ 0.010 x 10$^{11}$ years\\  
$^{252}$Cf& 86 $\pm$ 1 year\\  
$^{254}$Fm & 228 $\pm$ 1 day\\  
$^{258}$Fm & 0.37 $\pm$ 0.02 ms\\  
$^{256}$Rf & 6.2 $\pm$ 0.2 ms\\  
$^{260}$Rf & 20 $\pm$ 1 ms  
\end{tabular}  
\end{ruledtabular}  
\end{table}  
  
It is worth noting that when dealing with quantities that can vary by many orders of   
magnitude, it makes sense to compare not the differences between  
theory and experiment but rather the logarithm of the ratio of  
theory to experiment,   
\be  
R_x = \log\left({x_{th}\over x_{exp}}\right).  
\ee    
The target performance measures are then the mean value of $R_x$,  
\be  
\label{mean}  
\bar R_x  = {1\over N_d } \sum_i R_{x,i}  
\ee  
and the variance about the mean  
\be  
\label{variance}  
\sigma = {1\over N_d} \left(\sum_i (R_{x,i} -\bar R_x)^2\right)^{1/2}.  
\ee   
Here $N_d$ is the number of data points in the benchmark set. We note that these measures are in common use, for example in reporting the performance of theories of the nuclear   
level density~\cite{Hil06}.  Of course, if the model makes use of a parameter to fit benchmark data or data of the same kind, only the $\sigma$ value provides an interesting test of the theory.  
 
\subsection{Mass Distributions}  
 
Fission fragment yields are commonly characterized by {\it independent}, {\it cumulative} and {\it chain} mass yields. Establishing meaningful benchmarks is complicated by the fact that there is no direct relation between what theories predict and what experiments measure. 
 
Experimentally, the best-known mass yields are for the thermal neutron-induced fission reactions on $^{235}$U and $^{239}$Pu. Precise measurements (1$-$2\%) have often been made using radiochemical techniques, in which cumulative yields are measured. Inferring the independent yields from those measurements therefore requires some modeling. Finally, fission theories will predict {\it pre-neutron} emission fission fragment yields, while experimental data always correspond to {\it post-neutron} emission yields.

However, for benchmarking purposes, we just recommend only two quantities that should  
be easier to compute and reflect the coarsest features of the distribution. 

We first determine the average mass $A_m$ as
\be
A_m = {1\over P} \sum_A AP(A)
\ee
where $P = \sum_A P(A)$ is the total probability. Note that $P = 1$ is
not precisely satisfied in the evaluated data tables. The experimental
$A_m$ comes out a few units less than half the mass
number of the original nucleus.

The benchmarks are the following two moments of the distribution for the higher mass fragments:  
\begin{eqnarray}  
S_> = & {1\over P_>}\sum_{A>A_m}P(A) \left(A - A_m\right), \\  
\sigma_>^2 = & {1\over P_>}\sum_{A>A_m}P(A) \left(A - A_m\right)^2  
- S_>^2.  
\end{eqnarray}  
Here $P_>$ is the total probability of producing fission fragments of  
mass higher than $A_m$:  
\be  
P_> = \sum_{A>A_m} P(A).  
\ee    
In simple models $P_>$ will be equal to one, but the experimental  
value differs from that by a small amount.    
  
For the experimental cases, we include the thermal  
neutron-induced fission of $^{235}$U, $^{239}$Pu  and $^{255}$Fm.  
The first two have the classic asymmetric mass yields and the  
latter has a more centered yield curve.  Also we consider an example of spontaneous fission of $^{252}$Cf.  The moments  
in Table~\ref{fragments} were extracted from    
the experimental $P(A)$ data compiled in Refs.~\cite{Eng94,files}.  
The Table also gives the  
values of $A_m$ and $P_>$ for the data, although these are not part of the  
benchmark.  The full tables for $P(A)$ are provided in the Appendix.

\begin{table}[htb]   
\caption{\label{fragments}  
Fission product mass distribution characteristics  
extracted from the   
 experimental  data compiled in Ref.~\cite{Eng94}. (The data are available in a tabulated text form in Ref.~\cite{files}.)  
Asterisk denotes induced fission by thermal neutron capture  
on the $A-1$ isotope.}  
\begin{ruledtabular}  
\begin{tabular}{ccccc}   
Nuclide  & $A_m$ & $P_>$  & $S_>$  &  $\sigma_>$  \\  
\hline  \\[-8pt]  
$^{236}$U* & 116.7 & 0.98 & 22.0 &     5.1 \\  
$^{240}$Pu* & 118.3  & 0.96 &  19.9 & 5.7 \\  
$^{252}$Cf & 124.0 & 0.99 & 18.0 & 6.4 \\  
$^{256}$Fm* & 126.4 & 0.97 & 12.3 & 6.9   
\end{tabular}  
\end{ruledtabular}  
\end{table}  
  
\subsection{Total Kinetic Energies} 
 
The total kinetic energy ($TKE$) of the fission fragments is an important quantity for several reasons. It is an indicator for the shape of the fission fragments near their scission configurations: the higher the $TKE$ value, the more compact  the nascent fragments are. This quantity  also directly influences the excitation energy left in the initial fragments, which is  released through the evaporation of neutrons and photons. It also represents an important benchmark for fission theories to compute. 
 
The average pre-neutron evaporation total kinetic energies $\langle TKE\rangle$ for $^{252}$Cf spontaneous fission and thermal neutron-induced fission of $^{233,235}$U and $^{239}$Pu are considered as {\it energy standards}~\cite{Gonnenwein:1991}. To a first-order, the evolution of $\langle TKE\rangle$ follows the Coulomb parameter $Z^2/A^{1/3}$.

\begin{table}[htb] 
\caption{\label{tab:TKE}  
Recommended~\cite{Gonnenwein:1991} average pre-neutron evaporation total kinetic energies of the fission fragments.}  
\begin{ruledtabular} 
\begin{tabular}{cc} 
Reaction  & $\langle TKE \rangle$ (MeV) \\ 
\hline 
$^{233}$U ($n_{\rm th},f$) & 170.1 $\pm$ 0.5 \\ 
$^{235}$U ($n_{\rm th},f$) & 170.5 $\pm$ 0.5 \\ 
$^{239}$Pu ($n_{\rm th},f$) & 177.9 $\pm$ 0.5 \\ 
$^{252}$Cf (sf) & 184.1 $\pm$ 1.3 \\ 
\end{tabular}  
\end{ruledtabular}  
\end{table}

----------------------------------------------------------------- 
\section{Concluding Remarks} 
 
This document provides a small set of fission data that can be used to test the validity of theoretical calculations. Obviously the fission process is very complex and rich, and many more data exist beyond this very small sample. 
One should view these notes   as a living document, which will need to be updated as more useful information becomes available, and as fidelity of fission theory improves.

\bigskip 
\section{Acknowledgment}    
   These benchmarks arose out of the Program INT-13-3 at  the Institute for Nuclear Theory,  
``Quantitative Large Amplitude Shape Dynamics: fission and heavy ion fusion." Discussions with A. Andreyev, R. Mills, and A. Sonzogni are gratefully acknowledged. This work was supported by the U.S. Department of Energy under Contracts No. DE-FG02-00ER41132 (INT),  No. DE-SC0008511    (NUCLEI SciDAC Collaboration), No. DE-NA0002574 (Stewardship  
Science Academic Alliances program), and No. DE-FG06-97ER41026 (OSU).

\section{Appendix}  
  
This Appendix contains  the tabulated information  on individual  mass yield distributions  
for the cases listed in Table~\ref{fragments}.  
(From Ref.~\cite{Eng94} and {\tt ie.lbl.gov/fission.html}.)

{\small  
\begin{longtable}{ccc}  
\caption{Fission Product Yields per 100 Fissions for  
 $^{235}$U: thermal neutron induced fission.}\\  
\bfseries A &\bfseries  Chain Yield ($\%$) & \bfseries Average Z\\ \hline \hline    
\endfirsthead  
\bfseries A &\bfseries  Chain Yield ($\%$)& \bfseries Average Z  \\ \hline \hline    
\endhead  
\hline \multicolumn{3}{r}{\emph{Continued}}  
\endfoot  
\hline \hline   
\endlastfoot

66	&	7.22E-08	&	26.54	\\  
67	&	3.61E-07	&	27.06	\\  
68	&	7.16E-07	&	27.56	\\  
69	&	1.57E-06	&	27.85	\\  
70	&	3.62E-06	&	28.12	\\  
71	&	8.39E-06	&	28.49	\\  
72	&	2.65E-05	&	28.94	\\  
73	&	1.02E-04	&	29.41	\\  
74	&	3.39E-04	&	29.82	\\  
75	&	1.07E-03	&	30.09	\\  
76	&	3.10E-03	&	30.4	\\  
77	&	7.95E-03	&	30.68	\\  
78	&	2.09E-02	&	31.17	\\  
79	&	4.47E-02	&	31.6	\\  
80	&	1.28E-01	&	32.02	\\  
81	&	2.03E-01	&	32.34	\\  
82	&	3.25E-01	&	32.69	\\  
83	&	5.35E-01	&	33.31	\\  
84	&	8.93E-01	&	33.74	\\  
85	&	1.28E+00	&	34.13	\\  
86	&	1.94E+00	&	33.58	\\  
87	&	2.52E+00	&	34.85	\\  
88	&	3.53E+00	&	35.36	\\  
89	&	4.75E+00	&	35.81	\\  
90	&	5.89E+00	&	36.07	\\  
91	&	5.87E+00	&	36.43	\\  
92	&	5.97E+00	&	36.92	\\  
93	&	6.24E+00	&	37.37	\\  
94	&	6.58E+00	&	37.8	\\  
95	&	6.55E+00	&	38.09	\\  
96	&	6.02E+00	&	38.33	\\  
97	&	6.00E+00	&	38.89	\\  
98	&	5.76E+00	&	39.36	\\  
99	&	6.14E+00	&	39.72	\\  
100	&	6.30E+00	&	40.02	\\  
101	&	5.18E+00	&	40.39	\\  
102	&	4.30E+00	&	40.62	\\  
103	&	3.03E+00	&	41.23	\\  
104	&	1.88E+00	&	41.66	\\  
105	&	9.72E-01	&	41.67	\\  
106	&	4.02E-01	&	42.03	\\  
107	&	1.46E-01	&	42.14	\\  
108	&	5.41E-02	&	42.44	\\  
109	&	3.11E-02	&	42.53	\\  
110	&	2.55E-02	&	43.24	\\  
111	&	1.74E-02	&	43.77	\\  
112	&	1.30E-02	&	44.14	\\  
113	&	1.42E-02	&	44.64	\\  
114	&	1.18E-02	&	45.36	\\  
115	&	1.26E-02	&	45.8	\\  
116	&	1.32E-02	&	46.35	\\  
117	&	1.28E-02	&	46.28	\\  
118	&	1.14E-02	&	46.87	\\  
119	&	1.29E-02	&	47.41	\\  
120	&	1.26E-02	&	47.55	\\  
121	&	1.30E-02	&	48.07	\\  
122	&	1.55E-02	&	48.17	\\  
123	&	1.57E-02	&	48.39	\\  
124	&	2.68E-02	&	48.91	\\  
125	&	3.41E-02	&	49.4	\\  
126	&	5.83E-02	&	49.71	\\  
127	&	1.57E-01	&	49.64	\\  
128	&	3.48E-01	&	49.95	\\  
129	&	5.43E-01	&	50.02	\\  
130	&	1.81E+00	&	50.28	\\  
131	&	2.89E+00	&	50.79	\\  
132	&	4.31E+00	&	51.22	\\  
133	&	6.71E+00	&	51.65	\\  
134	&	7.84E+00	&	52.02	\\  
135	&	6.55E+00	&	52.5	\\  
136	&	3.90E+00	&	52.66	\\  
137	&	6.34E+00	&	53.44	\\  
138	&	6.76E+00	&	53.84	\\  
139	&	6.48E+00	&	54.1	\\  
140	&	6.76E+00	&	54.46	\\  
141	&	5.86E+00	&	55.07	\\  
142	&	5.83E+00	&	55.47	\\  
143	&	5.96E+00	&	55.82	\\  
144	&	5.51E+00	&	56.13	\\  
145	&	3.95E+00	&	56.51	\\  
146	&	3.00E+00	&	56.89	\\  
147	&	2.25E+00	&	57.66	\\  
148	&	1.68E+00	&	57.82	\\  
149	&	1.08E+00	&	58.21	\\  
150	&	6.53E-01	&	58.42	\\  
151	&	4.19E-01	&	58.95	\\  
152	&	2.67E-01	&	59.47	\\  
153	&	1.58E-01	&	59.8	\\  
154	&	7.44E-02	&	60.09	\\  
155	&	3.21E-02	&	60.45	\\  
156	&	1.48E-02	&	60.88	\\  
157	&	6.15E-03	&	61.38	\\  
158	&	3.29E-03	&	61.79	\\  
159	&	1.01E-03	&	62.05	\\  
160	&	3.19E-04	&	62.32	\\  
161	&	8.53E-05	&	62.79	\\  
162	&	1.59E-05	&	63.31	\\  
163	&	6.10E-06	&	63.67	\\  
164	&	1.88E-06	&	63.99	\\  
165	&	9.52E-07	&	64.29	\\  
166	&	3.62E-07	&	64.64	\\  
167	&	2.47E-07	&	65.16	\\  
168	&	5.70E-08	&	65.64	\\  
169	&	2.39E-08	&	65.92	\\  
170	&	5.01E-09	&	66.18	\\  
171	&	2.35E-09	&	66.58	\\  
172	&	7.69E-10	&	67.06	\\

\end{longtable}  
}

{\small  
\begin{longtable}{ccc}  
\caption{Fission Product Yields per 100 Fissions for  
 $^{239}$Pu: thermal neutron induced fission.}\\  
\bfseries A &\bfseries  Chain Yield ($\%$) & \bfseries Average Z\\ \hline \hline    
\endfirsthead  
\bfseries A &\bfseries  Chain Yield ($\%$)& \bfseries Average Z  \\ \hline \hline    
\endhead  
\hline \multicolumn{3}{r}{\emph{Continued}}  
\endfoot  
\hline \hline   
\endlastfoot  
		  
66	&	2.20E-07	&	27.16	\\  
67	&	4.49E-07	&	27.54	\\  
68	&	1.61E-06	&	27.90	\\  
69	&	5.89E-06	&	28.23	\\  
70	&	1.99E-05	&	28.57	\\  
71	&	3.67E-05	&	29.01	\\  
72	&	1.21E-04	&	29.45	\\  
73	&	2.62E-04	&	29.79	\\  
74	&	6.36E-04	&	30.11	\\  
75	&	1.37E-03	&	30.48	\\  
76	&	2.94E-03	&	30.87	\\  
77	&	7.23E-03	&	31.32	\\  
78	&	1.88E-02	&	31.71	\\  
79	&	4.37E-02	&	32.06	\\  
80	&	9.37E-02	&	32.37	\\  
81	&	1.84E-01	&	32.82	\\  
82	&	2.29E-01	&	33.26	\\  
83	&	2.96E-01	&	33.62	\\  
84	&	4.70E-01	&	34.01	\\  
85	&	5.81E-01	&	34.27	\\  
86	&	6.77E-01	&	34.55	\\  
87	&	9.90E-01	&	35.18	\\  
88	&	1.31E+00	&	35.63	\\  
89	&	1.72E+00	&	35.96	\\  
90	&	2.16E+00	&	36.34	\\  
91	&	2.49E+00	&	36.86	\\  
92	&	2.99E+00	&	37.27	\\  
93	&	3.75E+00	&	37.65	\\  
94	&	4.35E+00	&	38.00	\\  
95	&	4.85E+00	&	38.31	\\  
96	&	4.36E+00	&	38.56	\\  
97	&	5.41E+00	&	39.17	\\  
98	&	5.81E+00	&	39.51	\\  
99	&	6.23E+00	&	39.92	\\  
100	&	6.77E+00	&	40.21	\\  
101	&	6.02E+00	&	40.61	\\  
102	&	6.13E+00	&	41.11	\\  
103	&	6.99E+00	&	41.59	\\  
104	&	6.08E+00	&	41.89	\\  
105	&	5.65E+00	&	42.21	\\  
106	&	4.36E+00	&	42.56	\\  
107	&	3.33E+00	&	43.12	\\  
108	&	2.16E+00	&	43.53	\\  
109	&	1.48E+00	&	43.77	\\  
110	&	6.45E-01	&	43.99	\\  
111	&	2.96E-01	&	44.13	\\  
112	&	1.29E-01	&	44.27	\\  
113	&	8.17E-02	&	44.54	\\  
114	&	6.03E-02	&	44.93	\\  
115	&	4.26E-02	&	45.53	\\  
116	&	5.07E-02	&	46.02	\\  
117	&	4.45E-02	&	46.48	\\  
118	&	3.25E-02	&	46.85	\\  
119	&	3.23E-02	&	47.48	\\  
120	&	3.06E-02	&	47.91	\\  
121	&	3.78E-02	&	48.30	\\  
122	&	4.46E-02	&	48.72	\\  
123	&	4.41E-02	&	49.24	\\  
124	&	7.87E-02	&	49.55	\\  
125	&	1.12E-01	&	49.70	\\  
126	&	2.02E-01	&	49.87	\\  
127	&	5.07E-01	&	49.96	\\  
128	&	7.34E-01	&	50.07	\\  
129	&	1.37E+00	&	50.27	\\  
130	&	2.36E+00	&	50.66	\\  
131	&	3.86E+00	&	51.19	\\  
132	&	5.41E+00	&	51.43	\\  
133	&	7.02E+00	&	51.99	\\  
134	&	7.59E+00	&	52.34	\\  
135	&	7.63E+00	&	52.84	\\  
136	&	3.40E+00	&	52.85	\\  
137	&	6.71E+00	&	53.71	\\  
138	&	6.11E+00	&	53.94	\\  
139	&	5.66E+00	&	54.43	\\  
140	&	5.37E+00	&	54.97	\\  
141	&	5.25E+00	&	55.28	\\  
142	&	4.93E+00	&	55.72	\\  
143	&	4.42E+00	&	56.04	\\  
144	&	3.75E+00	&	56.37	\\  
145	&	2.99E+00	&	56.87	\\  
146	&	2.46E+00	&	57.34	\\  
147	&	2.01E+00	&	57.73	\\  
148	&	1.64E+00	&	58.32	\\  
149	&	1.22E+00	&	58.52	\\  
150	&	9.67E-01	&	58.88	\\  
151	&	7.38E-01	&	59.34	\\  
152	&	5.76E-01	&	59.74	\\  
153	&	3.61E-01	&	60.06	\\  
154	&	2.60E-01	&	60.38	\\  
155	&	1.66E-01	&	60.81	\\  
156	&	1.24E-01	&	61.25	\\  
157	&	7.42E-02	&	61.62	\\  
158	&	4.14E-02	&	61.97	\\  
159	&	2.06E-02	&	62.32	\\  
160	&	9.68E-03	&	62.66	\\  
161	&	4.85E-03	&	63.11	\\  
162	&	2.23E-03	&	63.54	\\  
163	&	9.17E-04	&	63.87	\\  
164	&	3.30E-04	&	64.18	\\  
165	&	1.34E-04	&	64.57	\\  
166	&	6.66E-05	&	64.98	\\  
167	&	1.52E-05	&	65.39	\\  
168	&	4.29E-06	&	65.78	\\  
169	&	1.47E-06	&	66.10	\\  
170	&	3.18E-07	&	66.42	\\  
171	&	1.57E-07	&	66.85	\\  
172	&	4.94E-08	&	67.29	\\  
  
\end{longtable}  
}

{\small  
\begin{longtable}{ccc}  
\caption{Fission Product Yields per 100 Fissions for  
 $^{252}$Cf: spontaneous fission.}\\  
\bfseries A &\bfseries  Chain Yield ($\%$) & \bfseries Average Z\\ \hline \hline    
\endfirsthead  
\bfseries A &\bfseries  Chain Yield ($\%$)& \bfseries Average Z  \\ \hline \hline    
\endhead  
\hline \multicolumn{3}{r}{\emph{Continued}}  
\endfoot  
\hline \hline   
\endlastfoot  
	  
66	&	5.26E-08	&	2.66E+01	\\  
67	&	1.44E-07	&	2.70E+01	\\  
68	&	3.84E-07	&	2.74E+01	\\  
69	&	9.90E-07	&	2.77E+01	\\  
70	&	2.39E-06	&	2.81E+01	\\  
71	&	6.02E-06	&	2.85E+01	\\  
72	&	1.41E-05	&	2.89E+01	\\  
73	&	3.22E-05	&	2.93E+01	\\  
74	&	7.06E-05	&	2.97E+01	\\  
75	&	1.52E-04	&	3.00E+01	\\  
76	&	3.17E-04	&	3.04E+01	\\  
77	&	6.25E-04	&	3.08E+01	\\  
78	&	2.06E-03	&	3.12E+01	\\  
79	&	3.44E-03	&	3.16E+01	\\  
80	&	4.71E-03	&	3.20E+01	\\  
81	&	8.36E-03	&	3.23E+01	\\  
82	&	1.52E-02	&	3.27E+01	\\  
83	&	4.13E-02	&	3.31E+01	\\  
84	&	5.24E-02	&	3.35E+01	\\  
85	&	1.22E-01	&	3.39E+01	\\  
86	&	1.18E-01	&	3.43E+01	\\  
87	&	2.08E-01	&	3.47E+01	\\  
88	&	3.06E-01	&	3.52E+01	\\  
89	&	3.60E-01	&	3.56E+01	\\  
90	&	5.43E-01	&	3.60E+01	\\  
91	&	6.00E-01	&	3.64E+01	\\  
92	&	6.78E-01	&	3.67E+01	\\  
93	&	8.82E-01	&	3.71E+01	\\  
94	&	1.11E+00	&	3.76E+01	\\  
95	&	1.25E+00	&	3.79E+01	\\  
96	&	1.56E+00	&	3.83E+01	\\  
97	&	1.67E+00	&	3.87E+01	\\  
98	&	2.27E+00	&	3.91E+01	\\  
99	&	2.65E+00	&	3.95E+01	\\  
100	&	3.46E+00	&	3.99E+01	\\  
101	&	3.93E+00	&	4.03E+01	\\  
102	&	4.04E+00	&	4.07E+01	\\  
103	&	5.45E+00	&	4.11E+01	\\  
104	&	5.64E+00	&	4.16E+01	\\  
105	&	6.23E+00	&	4.22E+01	\\  
106	&	6.32E+00	&	4.23E+01	\\  
107	&	6.62E+00	&	4.28E+01	\\  
108	&	6.10E+00	&	4.32E+01	\\  
109	&	5.94E+00	&	4.38E+01	\\  
110	&	5.91E+00	&	4.41E+01	\\  
111	&	5.19E+00	&	4.46E+01	\\  
112	&	4.13E+00	&	4.50E+01	\\  
113	&	4.78E+00	&	4.56E+01	\\  
114	&	3.33E+00	&	4.61E+01	\\  
115	&	2.90E+00	&	4.63E+01	\\  
116	&	2.13E+00	&	4.66E+01	\\  
117	&	1.50E+00	&	4.69E+01	\\  
118	&	9.94E-01	&	4.72E+01	\\  
119	&	3.84E-01	&	4.75E+01	\\  
120	&	2.39E-01	&	4.77E+01	\\  
121	&	1.18E-01	&	4.78E+01	\\  
122	&	8.84E-02	&	4.80E+01	\\  
123	&	4.09E-02	&	4.80E+01	\\  
124	&	2.54E-02	&	4.85E+01	\\  
125	&	1.77E-02	&	4.93E+01	\\  
126	&	2.77E-02	&	4.98E+01	\\  
127	&	1.06E-01	&	5.00E+01	\\  
128	&	1.93E-01	&	5.01E+01	\\  
129	&	5.88E-01	&	5.03E+01	\\  
130	&	8.47E-01	&	5.06E+01	\\  
131	&	1.60E+00	&	5.10E+01	\\  
132	&	2.15E+00	&	5.13E+01	\\  
133	&	3.15E+00	&	5.17E+01	\\  
134	&	3.86E+00	&	5.21E+01	\\  
135	&	4.19E+00	&	5.26E+01	\\  
136	&	3.23E+00	&	5.27E+01	\\  
137	&	5.09E+00	&	5.37E+01	\\  
138	&	5.56E+00	&	5.40E+01	\\  
139	&	5.89E+00	&	5.43E+01	\\  
140	&	5.96E+00	&	5.46E+01	\\  
141	&	5.97E+00	&	5.50E+01	\\  
142	&	6.02E+00	&	5.55E+01	\\  
143	&	6.25E+00	&	5.60E+01	\\  
144	&	5.89E+00	&	5.63E+01	\\  
145	&	5.07E+00	&	5.66E+01	\\  
146	&	4.44E+00	&	5.70E+01	\\  
147	&	4.28E+00	&	5.75E+01	\\  
148	&	3.94E+00	&	5.79E+01	\\  
149	&	2.73E+00	&	5.83E+01	\\  
150	&	2.44E+00	&	5.86E+01	\\  
151	&	1.95E+00	&	5.92E+01	\\  
152	&	1.72E+00	&	5.96E+01	\\  
153	&	1.29E+00	&	6.00E+01	\\  
154	&	1.07E+00	&	6.05E+01	\\  
155	&	7.92E-01	&	6.08E+01	\\  
156	&	6.76E-01	&	6.10E+01	\\  
157	&	5.38E-01	&	6.16E+01	\\  
158	&	4.70E-01	&	6.22E+01	\\  
159	&	3.40E-01	&	6.24E+01	\\  
160	&	2.86E-01	&	6.28E+01	\\  
161	&	1.94E-01	&	6.32E+01	\\  
162	&	1.20E-01	&	6.36E+01	\\  
163	&	7.58E-02	&	6.40E+01	\\  
164	&	4.72E-02	&	6.43E+01	\\  
165	&	2.87E-02	&	6.47E+01	\\  
166	&	1.84E-02	&	6.51E+01	\\  
167	&	9.57E-03	&	6.55E+01	\\  
168	&	5.25E-03	&	6.59E+01	\\  
169	&	1.67E-03	&	6.63E+01	\\  
170	&	1.40E-03	&	6.66E+01	\\  
171	&	7.09E-04	&	6.70E+01	\\  
172	&	3.46E-04	&	6.74E+01	\\  
  
\end{longtable}  
}

{\small  
\begin{longtable}{ccc}  
\caption{Fission Product Yields per 100 Fissions for  
 $^{255}$Fm: thermal neutron induced fission.}\\  
\bfseries A &\bfseries  Chain Yield ($\%$) & \bfseries Average Z\\ \hline \hline    
\endfirsthead  
\bfseries A &\bfseries  Chain Yield ($\%$)& \bfseries Average Z  \\ \hline \hline    
\endhead  
\hline \multicolumn{3}{r}{\emph{Continued}}  
\endfoot  
\hline \hline   
\endlastfoot  
66	&	1.59E-04	&	27.17	\\  
67	&	2.05E-04	&	27.55	\\  
68	&	2.52E-04	&	27.93	\\  
69	&	3.27E-04	&	28.31	\\  
70	&	4.38E-04	&	28.69	\\  
71	&	5.78E-04	&	29.07	\\  
72	&	7.46E-04	&	29.45	\\  
73	&	9.52E-04	&	29.83	\\  
74	&	1.31E-03	&	30.21	\\  
75	&	1.58E-03	&	30.59	\\  
76	&	2.15E-03	&	30.97	\\  
77	&	2.80E-03	&	31.35	\\  
78	&	3.72E-03	&	31.73	\\  
79	&	4.66E-03	&	32.11	\\  
80	&	6.34E-03	&	32.49	\\  
81	&	8.02E-03	&	32.86	\\  
82	&	1.03E-02	&	33.23	\\  
83	&	1.33E-02	&	33.63	\\  
84	&	1.62E-02	&	34.01	\\  
85	&	2.43E-02	&	34.41	\\  
86	&	2.64E-02	&	34.67	\\  
87	&	3.94E-02	&	35.19	\\  
88	&	5.11E-02	&	35.58	\\  
89	&	6.60E-02	&	35.98	\\  
90	&	8.66E-02	&	36.37	\\  
91	&	1.12E-01	&	36.77	\\  
92	&	1.39E-01	&	37.16	\\  
93	&	1.87E-01	&	37.56	\\  
94	&	2.36E-01	&	37.95	\\  
95	&	3.08E-01	&	38.34	\\  
96	&	3.78E-01	&	38.63	\\  
97	&	6.88E-01	&	39.13	\\  
98	&	6.97E-01	&	39.52	\\  
99	&	8.60E-01	&	39.92	\\  
100	&	9.52E-01	&	40.31	\\  
101	&	1.14E+00	&	40.7	         \\  
102	&	1.24E+00	&	41.1   	\\  
103	&	1.33E+00	&	41.49	\\  
104	&	1.52E+00	&	41.89	\\  
105	&	2.39E+00	&	42.28	\\  
106	&	2.41E+00	&	42.68	\\  
107	&	2.61E+00	&	43.07	\\  
108	&	2.71E+00	&	43.46	\\  
109	&	2.90E+00	&	43.86	\\  
110	&	3.28E+00	&	44.25	\\  
111	&	3.19E+00	&	44.64	\\  
112	&	3.63E+00	&	45.03	\\  
113	&	4.25E+00	&	45.41	\\  
114	&	4.84E+00	&	45.81	\\  
115	&	5.59E+00	&	46.2  	\\  
116	&	5.66E+00	&	46.58	\\  
117	&	5.85E+00	&	46.97	\\  
118	&	5.89E+00	&	47.35	\\  
119	&	5.97E+00	&	47.74	\\  
120	&	5.80E+00	&	48.14	\\  
121	&	5.70E+00	&	48.54	\\  
122	&	5.31E+00	&	49.01	\\  
123	&	4.92E+00	&	49.49	\\  
124	&	3.88E+00	&	49.8  	\\  
125	&	3.02E+00	&	49.91	\\  
126	&	2.41E+00	&	50	         \\  
127	&	2.32E+00	&	50.07	\\  
128	&	2.17E+00	&	50.13	\\  
129	&	2.27E+00	&	50.23	\\  
130	&	2.48E+00	&	50.46	\\  
131	&	3.21E+00	&	50.84	\\  
132	&	4.81E+00	&	51.24	\\  
133	&	5.42E+00	&	51.64	\\  
134	&	5.81E+00	&	52.06	\\  
135	&	6.19E+00	&	52.48	\\  
136	&	5.28E+00	&	52.73	\\  
137	&	6.47E+00	&	53.32	\\  
138	&	6.11E+00	&	53.73	\\  
139	&	5.72E+00	&	54.15	\\  
140	&	4.83E+00	&	54.58	\\  
141	&	4.62E+00	&	55.01	\\  
142	&	4.33E+00	&	55.45	\\  
143	&	3.14E+00	&	55.88	\\  
144	&	3.21E+00	&	56.3	         \\  
145	&	2.84E+00	&	56.71	\\  
146	&	2.48E+00	&	57.11	\\  
147	&	2.02E+00	&	57.52	\\  
148	&	1.74E+00	&	57.92	\\  
149	&	1.47E+00	&	58.31	\\  
150	&	1.29E+00	&	58.7	         \\  
151	&	1.29E+00	&	59.1	         \\  
152	&	1.10E+00	&	59.5   	\\  
153	&	9.98E-01	&	59.88	\\  
154	&	7.10E-01	&	60.27	\\  
155	&	5.31E-01	&	60.65	\\  
156	&	4.43E-01	&	61.04	\\  
157	&	3.65E-01	&	61.43	\\  
158	&	2.66E-01	&	61.81	\\  
159	&	1.95E-01	&	62.19	\\  
160	&	1.51E-01	&	62.57	\\  
161	&	1.15E-01	&	62.95	\\  
162	&	8.86E-02	&	63.34	\\  
163	&	7.09E-02	&	63.71	\\  
164	&	5.31E-02	&	64.09	\\  
165	&	4.59E-02	&	64.47	\\  
166	&	3.54E-02	&	64.85	\\  
167	&	2.66E-02	&	65.23	\\  
168	&	1.77E-02	&	65.61	\\  
169	&	1.51E-02	&	65.99	\\  
170	&	1.24E-02	&	66.37	\\  
171	&	8.86E-03	&	66.75	\\  
172	&	7.08E-03	&	67.13	\\  
\end{longtable}  
}

%

\end{document}